\documentclass[journal]{IEEEtran}
\usepackage{times}
\usepackage{epsfig}
\usepackage{graphicx}
\usepackage{amsmath}
\usepackage{amssymb}
\usepackage{multirow}
\usepackage{bm}
\usepackage{changepage}
\usepackage{float}
\usepackage{color}
\usepackage{mathrsfs}
\usepackage{array}

\usepackage{hyperref}
\hypersetup{hypertex=true,
	colorlinks=true,
	linkcolor=red,
	anchorcolor=green,
	citecolor=green}

\usepackage{fancyhdr}
\usepackage{lipsum}

\begin{document}

\title{Inter-slice Context Residual Learning for 3D Medical Image Segmentation}

\author{Jianpeng~Zhang,
	Yutong~Xie,~\IEEEmembership{Graduate Student Member, IEEE},\\
	Yan~Wang,
	and~Yong~Xia,~\IEEEmembership{Member,~IEEE}
	\thanks{This work was supported in part by the National Natural Science Foundation of China under Grants 61771397, and in part by the Science and Technology Innovation Committee of Shenzhen Municipality, China, under Grants JCYJ20180306171334997. Y. Xie was supported by the Innovation Foundation for Doctor Dissertation of Northwestern Polytechnical University under Grants CX202010. ({\em Corresponding author: Yong Xia})  }
	\thanks{J. Zhang, Y. Xie, Y. Wang, and Y. Xia are with the National Engineering Laboratory for Integrated Aero-Space-Ground-Ocean Big Data Application Technology, School of Computer Science and Engineering, Northwestern Polytechnical University, Xi'an 710072, China (e-mail: james.zhang@mail.nwpu.edu.cn; xuyongxie@mail.nwpu.edu.cn; yxia@nwpu.edu.cn).}
}

\markboth{IEEE Transactions on Medical Imaging}%
{Zhang \MakeLowercase{\textit{et al.}}: Context residual perceiving for 3D medical image segmentation}

\maketitle

\begin{abstract}
Automated and accurate 3D medical image segmentation plays an essential role in assisting medical professionals to evaluate disease progresses and make fast therapeutic schedules. 
Although deep convolutional neural networks (DCNNs) have widely applied to this task, the accuracy of these models still need to be further improved mainly due to their limited ability to 3D context perception. 
In this paper, we propose the 3D context residual network (ConResNet) for the accurate segmentation of 3D medical images. This model consists of an encoder, a segmentation decoder, and a context residual decoder. We design the context residual module and use it to bridge both decoders at each scale. Each context residual module contains both context residual mapping and context attention mapping, the formal aims to explicitly learn the inter-slice context information and the latter uses such context as a kind of attention to boost the segmentation accuracy.
We evaluated this model on the MICCAI 2018 Brain Tumor Segmentation (BraTS) dataset and NIH Pancreas Segmentation (Pancreas-CT) dataset. Our results not only demonstrate the effectiveness of the proposed 3D context residual learning scheme but also indicate that the proposed ConResNet is more accurate than six top-ranking methods in brain tumor segmentation and seven top-ranking methods in pancreas segmentation. Code is available at:   
  \def\UrlFont{\rm\small\ttfamily}
\url{https://git.io/ConResNet}

\end{abstract}

\begin{IEEEkeywords}
3D context perceiving, inter-slice context residual, 3D medical image segmentation.
\end{IEEEkeywords}

\IEEEpeerreviewmaketitle

\section{Introduction}

\begin{figure}[t]
	\begin{center}
		\includegraphics[width=1.0\linewidth]{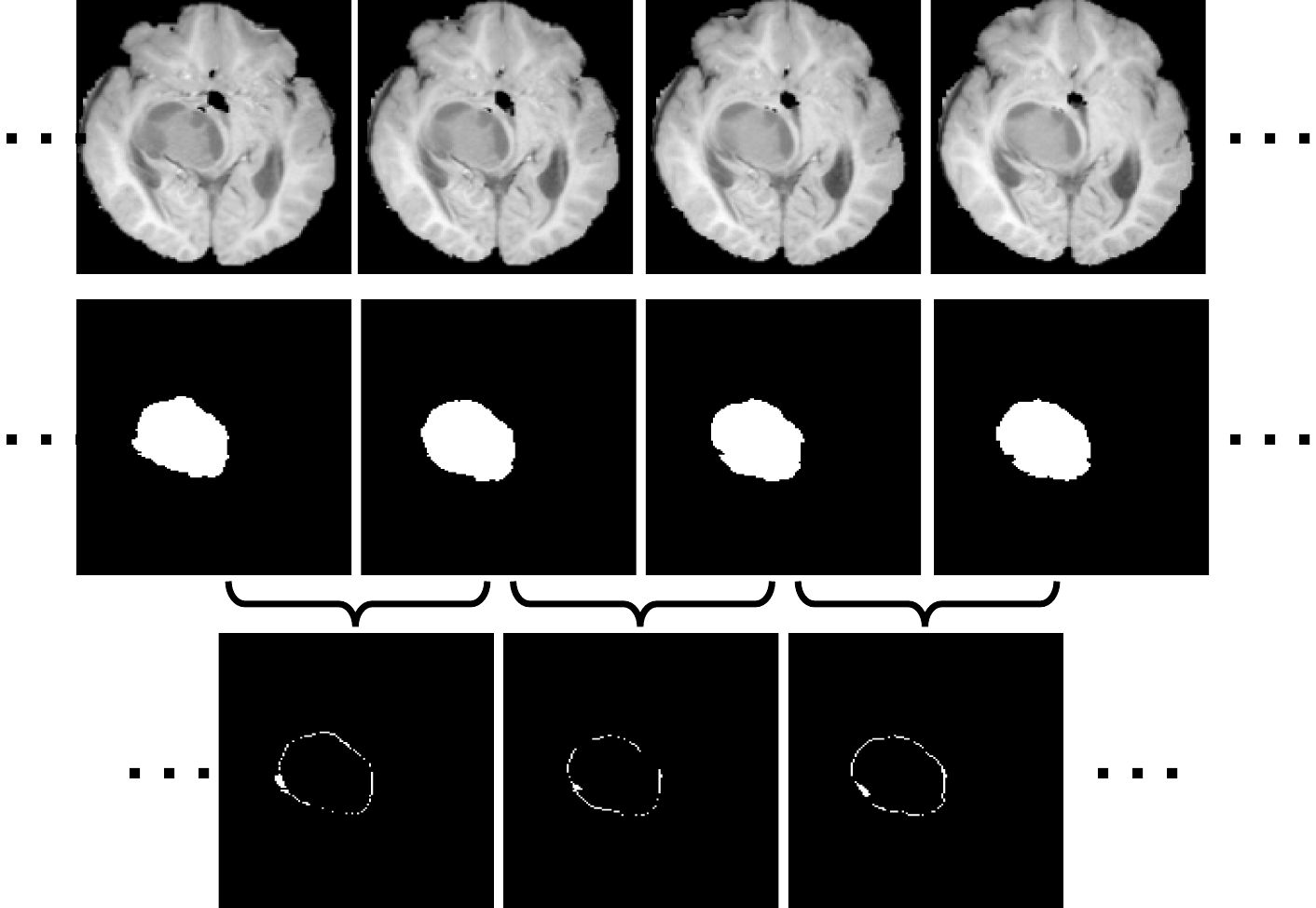}
	\end{center}
	\caption{Illustration of inter-slice context residual in 3D medical images. Top row: adjacent slices of a brain MRI image; middle row: corresponding masks of whole brain tumors; bottom row: inter-slice context residuals between adjacent slices (white: different; black: same). }
	\label{fig:problem}
\end{figure}

\IEEEPARstart{M}{edical} images provide visual representations of the anatomy or function of living bodies, which are essential for clinical analysis and medical intervention. Segmentation of 3D medical images aims to predict the semantic category (i.e. a specific organ or lesion) of each voxel, and is a fundamental and critical task in computer aided diagnosis (CAD), since accurate segmentation of organs or tumors is of value not only in facilitating the diagnosis but also in assessing the severity and prognosis of the disease.
This segmentation task, however, is extremely challenging due to the low soft tissue contrast and the heterogeneity of organs and tumors in shape, size, and location. Moreover, since the acquisition and annotation of medical data are expensive, there is usually a lack of sufficient annotated data to train segmentation models, which renders an even more challenging task for 3D medical image segmentation\cite{litjens2017survey,Xie_Mutual,Xie_SESV}. 

Recent years have witnessed the amazing success of deep convolutional neural networks (DCNNs) in image segmentation. Many attempts have been made to strengthen the ability of DCNNs to medical image segmentation. For instance, 
the encoder-decoder architecture has been improved in various ways to keep the low-level detailed information and obtain sharp object boundaries \cite{ronneberger2015unet,lin2017FPN,chen2018deeplabv3}, 
the spatial pyramid pooling has been used to exploit the multi-scale information \cite{chen2018deeplabv3}, 
the atrous convolution has been incorporated into segmentation models to expand the receptive field efficiently \cite{Fisher2016dilated}, 
and several attention learning mechanisms have been introduced to segmentation models, enabling them to focus more on certain locations and / or channels \cite{schlemper2019attentionUnet,sinha2019Attention}.
On 3D medical image segmentation tasks, 3D DCNNs have demonstrated striking improvements over their 2D counterparts \cite{dou20163d,dou20173d,chen2019med3d,zhou2019models}, since they are able to explore the contextual information contained across slices, which contributes significantly to better segmentation performance. We therefore advocate that the accuracy of 3D medical image segmentation can be further improved by capturing and using the inter-slice contextual information more effectively.

However, it is difficult to capture the inter-slice context information in volumetric images with complex anatomical structures. Take a brain tumor in a magnetic resonance (MR) sequence for example. In Fig.~\ref{fig:problem}, the first row shows four adjacent slices sampled from the sequence in which there is a brain tumor, the second row gives the ground truth tumor region in each slice, and the voxel-wise difference of the tumor region between any two adjacent slices, including extending outward regions or contracting inward regions, is displayed in the third row. We define such difference as the inter-slice context residual, which appears on or near the tumor surface and contains the essential and intriguing morphological information of the tumor, since we can use it, together with the tumor region in any slice, to reconstruct the shape of the 3D tumor. Intuitively, exploring the inter-slice context residual in a segmentation process must be beneficial to improving the accuracy. Unfortunately, since the tumor regions in two adjacent slices have the similar shape and size, the inter-slice context residual is usually tiny, and hence has never been characterized directly.

\begin{figure*}[t]
	\begin{center}
		\includegraphics[width=1.0\linewidth]{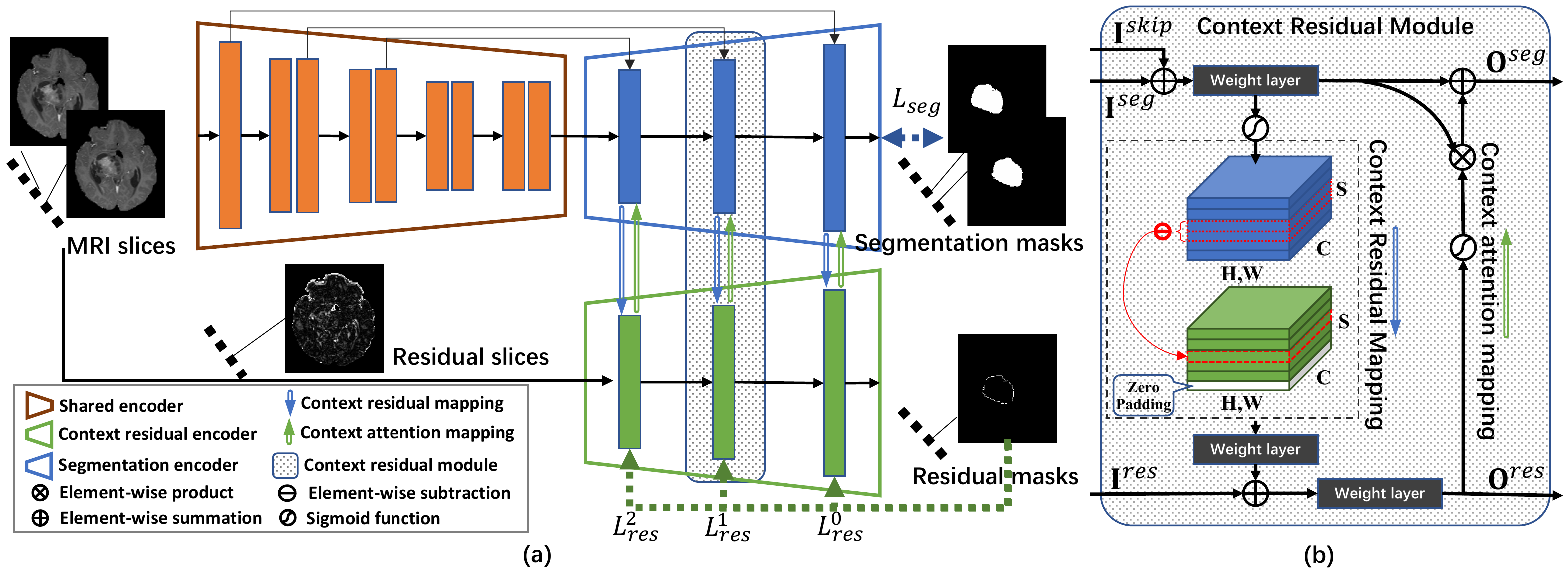}
	\end{center}
	\caption{(a) Diagram of the proposed ConResNet model, which has an encoder-decoder architecture, including a shared encoder (orange), a segmentation decoder (blue), and a ConRes decoder (green). (b) Context residual module.}
	\label{fig:framework}
\end{figure*}

In this paper, we propose the 3D context residual network (ConResNet) for accurate segmentation of 3D medical images. ConResNet has an encoder-decoder architecture, containing an encoder for feature extraction and two decoders for the generation of segmentation masks and inter-slice context residuals, respectively. The context residual (ConRes) decoder takes the residual feature maps of adjacent slices produced by the segmentation decoder as its input, and also provides feedback to the segmentation decoder as a kind of attention guidance, aiming to boost the ability of the segmentation model to perceive and use the inter-slice context information effectively (see Fig. ~\ref{fig:framework}). The design of ConRes decoder is conceptually generic and compatible with any existing 3D DCNN-based medical image segmentation model. We evaluate the proposed ConResNet model on the MICCAI 2018 Brain Tumor Segmentation (BraTS) dataset and NIH pancreas segmentation dataset and achieve the state-of-the-art performance on both segmentation tasks. The main contributions of this paper are summarized as follows:
\begin{itemize}
	\item We suggest adding the ConRes decoder to an encoder-decoder structure to explicitly boost the model's ability to 3D context perception and thus improve the segmentation accuracy.
	\item We design the context residual module, which is used between the segmentation decoder and ConRes decoder at each scale, for simultaneous context residual mapping and context attention mapping.
	\item We propose an accurate 3D medical image segmentation model called ConResNet, which achieves improved performance over state-of-the-art methods on both brain tumor segmentation task and pancreas segmentation task. 
\end{itemize}

\section{Related work}

\subsection{Context learning} 
The context of an object provides the information of its surroundings, and hence is essential for semantic segmentation. Many strategies have been developed for context learning, which can be roughly grouped into three categories.
First, to explore the context information at multiple scales, Zhao et al. \cite{zhao2017PSP} employed the pyramid spatial pooling strategy at different scales to aggregate multi-scale global information. Chen et al. \cite{chen2018deeplabv3} introduced the atrous spatial pyramid pooling with different dilated rates to parallel branches, which perform multi-scale representation aggregation. 
Second, to expand the receptive field, Yu et al. \cite{Fisher2016dilated} replaced the conventional convolution by the atrous convolution with an adjustable dilated rate, which shows superior performance on many computer vision tasks, like segmentation \cite{wei2018revisiting}, detection\cite{song2018pyramid}, and super-resolution\cite{zhang2018dcsr}.
Peng et al. \cite{peng2017largeKernel} utilized large kernels to capture rich global context information, which is beneficial for dense per-pixel prediction tasks.
Third, many attention based methods were proposed to filter out the extraneous information. Hu et al. \cite{hu2018squeeze} recalibrated adaptively channel-wise feature responses by explicitly modeling the channel-wise interdependencies of convolutional features. Wang et al. \cite{wang2017residual} designed an attention module with a bottom-up and top-down feed-forward structure to learn soft attention weights. In our previous work \cite{zhang2019attention}, we exploited the intrinsic self-attention ability of deep networks by using the high-level features to generate the attention map for low-level layers. 

These strategies have demonstrated their effectiveness in many 2D scenarios, and some of them have been extended to 3D cases. 
Wang et al. \cite{wang2019deeply} utilized a 3D fully convolutional networks (FCN) with group dilated convolutions to segment the prostate in MR sequences.
Schlemper et al. \cite{schlemper2019attentionUnetUnet} explored a 3D attention gated network which highlights task-related salient regions by embedding the attention mechanism into a U-like segmentation network for 3D organ segmentation. 
We also attempted to use the 3D atrous spatial pyramid pooling to capture multi-scale information for liver tumor segmentation in CT volumes \cite{zhang2019light}. 


Although improving the performance, these 2D and 3D strategies fail to characterize directly the inter-slice context residual information, which is particularly important for 3D medical image segmentation. 

\subsection{Residual learning} 
Residual learning can be traced back to the seminal work by He et al. \cite{he2016res}, Since then, it has been successfully applied to many computer vision tasks, including classification, segmentation, detection, and tracking. The idea of residual learning is to learn the residual between the input and output feature maps using the skip connection that directly jumps from input to output.
With the skip connection, residual learning eliminates the degradation problem, and hence makes it possible to train very deep networks with up to hundreds, even thousands, of layers.
In this work, the proposed ConRes decoder targets at perceiving the feature residual between two adjacent slices, and thus strengthens the model's ability to use the 3D context information for segmentation. 


\subsection{Medical image segmentation} 
Recently, both 2D and 3D DCNNs have become increasingly prevalent tools for medical image segmentation \cite{yu2016automated,wang2019abdominal,wang2018training,zhang2019light,Jia_APA,Li_Reinforce}.
Ronneberger et al. \cite{ronneberger2015unet} proposed a U-like architecture (UNet) that consists of an encoder path to capture segmentation-related high-level semantics and a symmetric decoder with skip connections from the encoder to generate segmentation results, and achieved excellent performance on several 2D medical image segmentation tasks. 
Fu et al. \cite{fu2018joint} designed a 2D M-Net, which combines a multi-scale U-like network with a side-output layer, and boosted the performance of optic disc and cup segmentation.
Chen et al. \cite{chen2016dcan} utilized a 2D deep contour-aware network to harness the multi-level contextual feature representation in an end-to-end way for effective gland segmentation.
When applied to 3D medical images, these 2D models perform the segmentation task in a slice-by-slice manner, and hence cannot capture inter-slice context information, leading to limited segmentation accuracy \cite{vorontsov2018liver}. 

With 3D convolutions, 3D DCNNs can directly process volumetric data and thus have distinct advantages over 2D DCNNs in 3D medical image segmentation.
Dou et al. \cite{dou20173d} designed a 3D fully convolutional network to generate high-quality score maps for automated 3D liver segmentation.
Li et al. \cite{li20183d} incorporated a multi-scale context module into a 3D segmentation network to use the multi-scale contextual information for inter-vertebral discs segmentation. 
Meanwhile, Chen et al. \cite{yalin2019ACloss} integrated the contour length and region constraints into the loss function to enforce the smoothness of segmented regions.
Karimi and Salcudean \cite{karimi2019hausdorff} introduced a Hausdorff distance based loss function to minimise the maximum deviation between the prediction and ground truth surfaces of targets.
Previously, we developed a light-weight hybrid convolutional network which replaces the 3D convolutions at bottom of network with low-cost 2D convolutions to reduce the model parameters and improve the segmentation performance with only the limited training data \cite{zhang2019light}.

Instead of designing a new segmentation model, we propose to improve the accuracy of existing 3D DCNNs via adding the ConRes decoder to them, which is able to capture and use the inter-slice context residual information.

\section{Method}

Let a 3D medical image be denoted by $\textbf{X} \in \textbf{R}^{S\times H \times W}$, where $S$ is the number of slices, and $H$ and $W$ are height and width of each slice, respectively. Its ground-truth segmentation mask is denoted by $\textbf{Y}_{seg} \in \textbf{R}^{S\times H \times W}$, in which the semantic label of each voxel is $\textbf{Y}_{seg}^{s, h, w} = \{0: background, 1: object\}$. Its ground-truth context residual mask is denoted by $\textbf{Y}_{res} \in \textbf{R}^{S\times H \times W}$, in which each element is calculated as 
\begin{equation}
\textbf{Y}_{res}^{s+1, h, w} = \left |  \textbf{Y}_{seg}^{s+1, h, w} - \textbf{Y}_{seg}^{s, h, w} \right |
\end{equation}
where $\textbf{Y}_{res}^{s+1, h, w}=0$ means that the voxels at $(h,w)$ on two adjacent ($s^{th}$ and $s+1^{th}$) slices belong to the same semantic categories, i.e., both of them are background or foreground, and $\textbf{Y}_{res}^{s+1, h, w}=1$ means that one voxel is background and the other is foreground. 

The proposed ConResNet aims to predict the segmentation mask $\textbf{Y}_{seg}$ and residual mask $\textbf{Y}_{res}$ simultaneously, formally shown as follows
\begin{equation}
\textbf{P}_{seg}, \textbf{P}_{res} = f(\textbf{X};\bm{\theta})
\end{equation}
where $\textbf{P}_{seg} \in \textbf{R}^{S\times H \times W}$ is the segmentation prediction of $\textbf{X}$, $\textbf{P}_{res} \in \textbf{R}^{S\times H \times W}$ is the residual prediction of $\textbf{X}$, and $\bm{\theta}$ represents the parameters of ConResNet. Accordingly, ConResNet has a shared encoder for feature extraction and two decoders for the prediction of $\textbf{Y}_{seg}$ and $\textbf{Y}_{res}$, respectively. The context residual module is designed to bridge both decoders, which is embedded in each pair of layers of dual decoders. The semantic features from the segmentation decoder are transformed to the context residual features through a new defined operation, called context residual mapping, and input to the ConRes decoder for the refinement. Besides, the ConRes decoder, in turn, provides an attention guidance via the context attention mapping to strengthen the 3D context perceiving ability of the segmentation decoder. The diagram of ConResNet is illustrated in Fig.~\ref{fig:framework}. We now delve into the details of this model.

\subsection{Shared Encoder}
In the proposed ConResNet, the shared encoder consists of nine residual blocks, each consisting of two $3\times3\times3$ convolutional layers and a skip connection from input to output. Due to the limited GPU memory, we have to train the network with very small batch size, like one sample per GPU. Therefore, to speed up the training process, we use the group normalization \cite{wu2018group} with a group number of eight, which is insensitive to the scale of batch size, to adjust and scale the activation of each layer.
Besides, we use the weight standardization algorithm \cite{weightstandardization} to accelerate micro-batch training by normalizing the weights of convolutional layers.
As shown in Fig.~\ref{fig:framework}, the encoding process can be divided into five stages.
In the first stage, the input is processed by a convolutional layer with 32 kernels and a residual block.
In each of next three stages, the data is processed by a convolutional layer with doubled kernels and a stride of 2 and two residual blocks. Thus, we gradually down-sample the feature maps to 1/8 of the input size and, simultaneously, increase the number of channels from 32 to 256, leading to the expansion of receptive field and reduction of computation. In the last stage, we employ two residual blocks, which use the atrous convolution with a dilated rate of 2, to further expand the receptive field while keeping the feature resolution for more details of the shape and edge.

\subsection{Dual Decoders}
Our ConResNet contains two decoders, i.e., a segmentation decoder and a ConRes decoder. The generation of a segmentation masks by the segmentation decoder or a residual mask by the ConRes decoder consists of three stages.
In each stage, we propose a context residual module to bridge the segmentation decoder and ConRes decoder. Specifically, the segmentation decoder first uses the trilinear interpolation to up-sample the previous feature maps, and then fuses them with the low-level features passing from the encoder using element-wise summation. The context residual of fused features are transmitted to the ConRes decoder. The ConRes decoder fuses the context residual features passing from the segmentation decoder and the features passing from the previous layer, and then refine them to predict the residual mask. Moreover, the inter-slice context residual information generated by the ConRes decoder is transmitted back to the segmentation decoder as an attention guidance to boost the ability of the segmentation decoder to perceive 3D context.
To match the number of channels in the encoder, we keep halving the channels after each up-sampling operation. 
As a result, we obtain a segmentation mask and the corresponding inter-slice context residual mask for each input volume.

\begin{figure}[t]
	\begin{center}
		\includegraphics[width=1.0\linewidth]{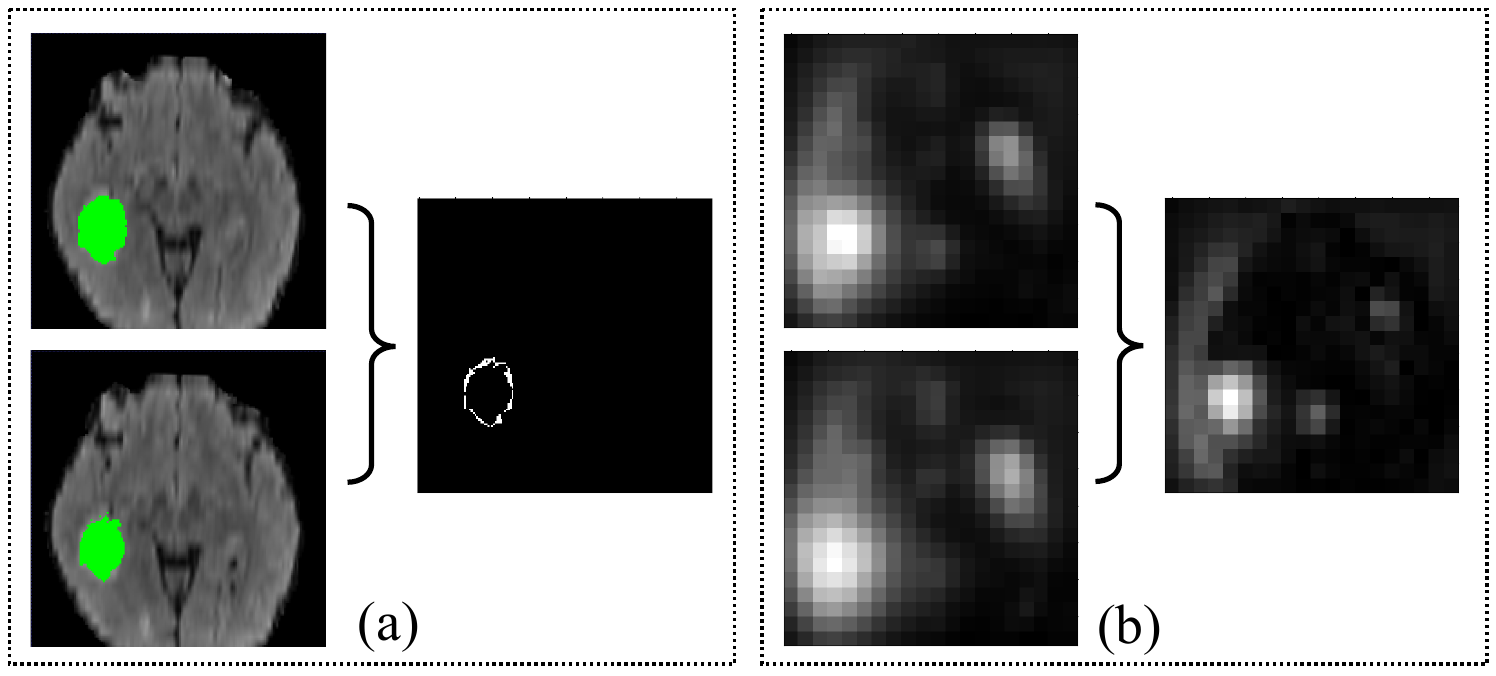}
	\end{center}
	\caption{Illustration of the context residual in images or feature maps. (a) Context residual of brain tumors (green) existing in two adjacent MR slices. (b) Context residual existing in the feature maps (extracted from output of encoder), where the left column gives the channel-wise average of the two adjacent feature maps along the inter-slice dimension. Note that each intensity value in the residual map indicates the level of dissimilarity between two voxels located at the same position in adjacent slices.}
	\label{fig:res}
\end{figure}

\subsection{Context Residual Module}
Each object in a 3D medical image, such as an organ or a tumor, gradually changes cross-sectional areas on 2D slices. As highlighted by those white pixels in Fig.~\ref{fig:res}(a), the context residual indicates the discrepancy between the cross-sectional areas on two adjacent slices, which is usually tiny. We suspect that such discrepancy can also be found in the feature maps of adjacent slices produced the segmentation decoder. To verify this, we visualize the channel-wise average of the feature maps of two adjacent slices and their difference in Fig.~\ref{fig:res}(b)). It shows that those two average feature maps look similar in most regions and the dissimilarity only appear in a small fraction of the slice, which is similar to the pattern of context residual shown in Fig.~\ref{fig:res}(a). Therefore, we design a context residual module to capture the context residual based on the feature maps produced the segmentation decoder.

Fig.~\ref{fig:framework}(b) shows the diagram of the context residual module. It is composed of two main paths, i.e., a segmentation path (top) and a context residual path (down).
In the segmentation path, the input is the element-wise summation of the segmentation features $\textbf{I}_{seg}$ coming from the previous layer and the low-level features $\textbf{I}_{skip}$ passing from the same-scale layer in the encoder, and the output is the segmentation feature $\textbf{O}_{seg}$. 
In the context residual path, the input is the context residual features $\textbf{I}_{res}$ coming from the previous layer and the output is the context residual features $\textbf{O}_{res}$. 
We design two strategies, i.e., context residual mapping and context attention mapping, to connect two paths for better context perceiving and semantic segmentation.

\noindent\textbf{Context residual mapping:} 
We use a weighted layer to refine the element-wise summation of $\textbf{I}_{seg}$ and $\textbf{I}_{skip}$, formally shown as
\begin{equation}
\textbf{F} = {\rm Conv}( \textbf{I}_{seg} \oplus  \textbf{I}_{skip}), 
\end{equation}
where $\oplus$ represents the element-wise summation, ${\rm Conv}$ represents the weighted layer, including convolution, group normalization, and ReLU activation.
The obtained feature map $\textbf{F}$ has a size of $S' \times H' \times W'$, where $S', H', W'$ represents the inter-slice depth, spatial height and width, respectively. 
The context residual $\textbf{G}$ is calculated as the position-wise absolute difference between each pair of adjacent features along the inter-slice dimension $S$, shown as follows
\begin{equation}
\textbf{G}^{s+1, h,w} = |\sigma (\textbf{F}^{s+1, h,w}) - \sigma (\textbf{F}^{s, h,w})|,
\end{equation}
where $\sigma(\cdot )$ is the sigmoid function.
Note that $\textbf{G}$ has a size of $(S'-1)\times H' \times W'$. For the convenience of subsequent processing, we pad the first-slice of $\textbf{G}$ to make it the same size of $\textbf{F}$.
Then, we combine the generated context residual feature map $\textbf{G}$ with $\textbf{I}_{res}$ from the previous context residual layer as the output of context residual path, shown as follows
\begin{equation}
\textbf{O}_{res} = {\rm Conv}({\rm Conv}(\textbf{G}) \oplus \textbf{I}_{res}).
\end{equation}

\noindent\textbf{Context attention mapping:} The context residual mapping generates the inter-slice context residual feature map and feeds it to the residual path for the context residual learning. In the meantime, the output of context residual path can be used as a kind of attention to boost the context perceiving ability of the segmentation path. To this end, we introduce the simple but effective context attention mapping.

We apply the sigmoid function to the context residual feature map $\textbf{O}_{res} $ and obtain the context attention weights, which is employed as an attention gate to activate the context residual regions and prompt the sensitivity of 3D context variance. Since the context attention highlights only the residual region (see Fig.~\ref{fig:framework}), directly applying the context attention weights to the feature maps may attenuate both the background and foreground. Hence, we define the context attention weighted output of the segmentation path as
\begin{equation}
\textbf{O}_{seg} = \textbf{F} \otimes (1+\sigma (\textbf{O}_{res}))
\end{equation}
where $\otimes$ means the element-wise multiplication.

\subsection{Network Optimization} 

For the segmentation decoder, the loss function is defined as the following combination of the cross entropy loss and Dice loss
\begin{equation}
\small
L_{seg} = \sum_{s,h,w} l_{bce}(\textbf{P}_{seg}^{s,h,w}, \textbf{Y}_{seg}^{s,h,w}) - \frac{2 \sum_{s,h,w} \textbf{P}_{seg}^{s,h,w} \textbf{Y}_{seg}^{s,h,w}}{\sum_{s,h,w}(\textbf{P}_{seg}^{s,h,w} + \textbf{Y}_{seg}^{s,h,w}) + \epsilon }
\end{equation}
where $l_{bce}$ is the binary cross entropy loss, $\textbf{P}_{seg}^{shw}$ and $\textbf{Y}_{seg}^{shw}$ are the predicted and ground-truth segmentation label at $(s, h, w)$, respectively, and $\epsilon$ is a smooth factor in the Dice loss. 

Training the ConRes decoder is supervised by the ground-truth context residual mask $\textbf{Y}^{res}$. The deep supervision technique \cite{dou20173d} (see Fig.~\ref{fig:framework}) is adopted to accelerate the convergence. Note that an additional convolutional layer is employed to predict the residual masks for each deep supervision operation.
Since the foreground and background voxels are highly imbalanced in each context residual mask, we set a weight for foreground voxels in the binary cross entropy loss function, shown as follows 
\begin{equation}
L_{res}^{(i)}  = \sum_{s,h,w} w_k l_{bce}({\textbf{P}_{res}^{s,h,w}}^{(i)}, \textbf{Y}_{res}^{s,h,w})
\end{equation}
where $i$ ($=0, 1, 2$) represents $i$-th context residual prediction, the class weight $w_k = \log \frac{V}{V_k}$, and $V_k$ is the number of voxels belonging to class $k$. 
The overall loss for context residual supervision is defined as
\begin{equation}
L_{res} = L_{res}^{0} + \lambda (L_{res}^{1} + L_{res}^{2})
\end{equation}
where the trade-off parameter $\lambda$ controls the influence of the loss in front layers and is empirically set to 0.5.

Hence, the proposed ConResNet can be jointly optimized by the core segmentation and auxiliary context residual supervisions via minimizing the joint loss $L = L_{seg} + L_{res} $ in an end-to-end manner.

\section{Experiments}

\begin{table*}[t]
\centering
\caption{Performance of the proposed ConResNet and six state-of-the-art segmentation methods on the BraTS 2018 online test dataset.}
\begin{tabular}{c|c|c|c|c|c|c|c|c}
\hline
\multirow{2}{*}{Methods}  & \multirow{2}{*}{Additional data} & \multicolumn{3}{c|}{Dice score (\%) / ranking} & \multicolumn{3}{c|}{Hausdorff distance / ranking} & \multirow{2}{*}{Sum-Score} \\ \cline{3-8}
& & ET & WT & TC & ET & WT & TC & \\ \hline \hline
CascadeNet \cite{wang2018automatic} & N & 79.72 / 7 & 90.21 / 8 & 85.83 / 5 & 3.13 / 6 & 6.18 / 8 & 6.37 / 5 & 39 \\ \hline
DMFNet \cite{chen2019dmfnet} & N & 80.12 / 6 & 90.62 / 5 & 84.54 / 8 & 3.06 / 5 & 4.66 / 6 & 6.44 / 3 & 33 \\ \hline
OM-Net \cite{zhou2020oneTIP} & N & 81.11 / 4 & 90.78 / 4 & 85.75 / 6 & 2.88 / 4 & 4.88 / 7 & 6.93 / 8 & 33 \\ \hline
DeepSCAN \cite{mckinley2018ensembles} & N & 79.60 / 8 & 90.30 / 7 & 84.70 / 7 & 3.55 / 7 & 4.17 / 2 & 4.93 / 1 & 32 \\ \hline
VAE-Seg \cite{myronenko20183d} & N & 82.33 / 2 & 91.00 / 2 & 86.68 / 1 & 3.93 / 8 & 4.52 / 5 & 6.85 / 7 & 25 \\ \hline
EnsembleNets \cite{zhou2018learning} & N & 81.36 / 3 & 90.95 / 3 & 86.51 / 2 & 2.72 / 2 & 4.17 / 2 & 6.54 / 6 & 18 \\ \hline
nnUnet \cite{isensee2018no} & Y & 80.87 / 5 & 91.26 / 1 & 86.34 / 3 & 2.41 / 1 & 4.27 / 4 & 6.52 / 4 & 18 \\ \hline
ConResNet & N & 83.15 / 1 & 90.38 / 6 & 85.90 / 4 & 2.79 / 3 & 4.15 / 1 & 5.75 / 2 & \textbf{17} \\ \hline
\end{tabular}
\label{tab:brats}
\end{table*}

\begin{table}[!t]
	\caption{Comparison of our method and the state-of-the-art methods on the Pancreas-CT dataset. }
	\begin{center}
		\begin{tabular}{c|c|c|c}
			\hline
			Methods     & Mean Dice & Max Dice & Min Dice \\ \hline \hline
			HNN \cite{roth2018spatial} & 81.27           & 88.96          & 50.69          \\ \hline
			Attention-UNet \cite{schlemper2019attentionUnet} & 83.10           & -          & -          \\ \hline
			CNN-RNN \cite{cai2018pancreas} & 83.70           & 91.00          & 59.00          \\ \hline
			RSTN \cite{yu2018recurrent} & 84.50           & 91.02          & 62.81          \\ \hline
			ResDSN \cite{zhu20173d}  & 84.59           & 91.45          & 69.62          \\ \hline
			Bayesian \cite{ma2018novel}  & 85.32           & 91.47          & 71.04          \\ \hline
			Coarse2Fine \cite{zhao2019fully} & 85.99           & 91.20          & 57.20          \\ \hline
			Our baseline & 85.23 & 91.90 & 67.40 \\ \hline
			ConResNet        & \textbf{86.06}                &  \textbf{92.00}             & \textbf{73.40}              \\ \hline
		\end{tabular}
	\end{center}
	\label{tab:PancreasCT}
\end{table}

\subsection{Datasets}
Two publicly available datasets were used for this study.

\noindent{\textbf{BraTS Dataset:}}
	The BraTS dataset \cite{BraTS_TMI} was collected and shared by the MICCAI 2018 Brain Tumor Segmentation Challenge \cite{BraTS_TMI}. The aim of this challenge is to develop automated segmentation algorithms to delineate intrinsically heterogeneous brain tumors, i.e., (1) enhancing tumor (ET), (2) tumor core (TC) that consists of ET, necrotic and non-enhancing tumor core, and (3) whole tumor (WT) that contains TC and the peritumoral edema. The BraTS dataset has 285 cases for training and 66 cases for online testing. 
	Each case contains four MR sequences, including the T1, T1c, T2, and FLAIR. All sequences were registered to the same anatomical template and interpolated to the same size of $155\times240\times240$ voxels and the same voxel size of $1.0\times1.0\times1.0$ $mm^3$. 
	The voxel-wise segmentation ground truth of training cases are publicly available, but the ground truth of validation cases are withheld for online evaluation. 

\noindent{\textbf{Pancreas-CT Dataset:}}
	Pancreas-CT\footnote{\url{https://wiki.cancerimagingarchive.net/display/Public/Pancreas-CT\#82af2dca8f2443b1bef1e85ac73acd44}} is the most authoritative open source dataset for pancreas segmentation, which is collected by the National Institutes of Health Clinical Center \cite{roth2015deeporgan}. 
	This dataset contains 82 contrasted-enhanced abdominal 3D CT scans from 53 male and 27 female subjects. The size of each scan is $512\times512\times(181\sim466)$ voxels and their voxel spacing varies from 0.5mm to 1.0mm. 
	Following \cite{roth2018spatial,cai2018pancreas,yu2018recurrent,zhu20173d,ma2018novel,zhao2019fully}, we adopted the 4-fold cross-validation and split randomly the dataset into four fixed and roughly equal folds. The average segmentation performance over all 82 cases was reported.

\subsection{Implementation Details}
	We implemented our ConResNet using PyTorch and performed all experiments on a workstation with two NVIDIA 2080Ti GPUs. During the training, we optimized ConResNet using the Adam algorithm with a batch size of 2 and weight decay of 0.0005. We set the initial learning rate to 0.0001 and decayed it according to a polynomial schedule 
	\begin{equation}
	lr^{t}=lr^{0}\times(1-t/T)^{0.9}
	\end{equation}
	where $t$ is the index of current iteration, and $T$ is the total number of iterations. To alleviate the overfitting, we employed simple online data augmentation techniques, including randomly scaling and flipping along three dimensions. 
	
	On the BraTS dataset, we first normalized the voxel values in each MR sequence to the standard normal distribution, and then concatenated four available modalities into a multimodal volume with 4 channels \cite{myronenko20183d}. In the training stage, we randomly cropped sub-volumes of size $80\times160\times160$ as training samples. 
	We treated this multi-class segmentation problem as three binary segmentation tasks. Accordingly, the last decoding layer of our ConResNet has three output channels, which use the sigmoid activation to produce the segmentation results of ET, WT, and TC, respectively. In this way, we optimized three sub-regions (\textit{i.e.}, ET, WT, and TC) directly. 
	
	On the Pancreas-CT dataset, we truncated the HU values of all voxels to the range [-100, +240] to remove irrelevant information \cite{zhu20173d}, and then linearly mapped HU values to [0, 1]. We randomly cropped $64\times120\times120$ patches as training samples. 

\noindent{\textbf{Evaluation Metrics:}}
	For this study, we adopted the Dice coefficient and Hausdorff distance to measure the performance of all segmentation methods on the BraTS dataset \cite{wang2018automatic,mckinley2018ensembles,myronenko20183d,zhou2018learning,isensee2018no}, and adopted the mean Dice score, max Dice score, and min Dice score as golden performance indicators on the Pancreas-CT dataset \cite{roth2018spatial,cai2018pancreas,yu2018recurrent,zhu20173d,ma2018novel,zhao2019fully}.
	
	The Dice coefficient is a statistic used to gauge the overlapping between a segmentation prediction $\textbf{P}_{seg}$ and the corresponding ground truth $\textbf{Y}_{seg}$. Specifically, the Dice score is defined as 
	\begin{equation}
	Dice = 2 \times \frac{|\textbf{P}_{seg} \cap  \textbf{Y}_{seg}|}{|\textbf{P}_{seg}| + |\textbf{Y}_{seg} | + \epsilon}
	\end{equation}
	The Hausdorff distance evaluates the quality of segmentation boundaries by computing the maximum distance between the prediction and its ground truth, defined as follows
	\begin{equation}
	\begin{aligned}
	HD = \max \{ &\sup_{p\in \partial\textbf{P}_{seg}} \inf_{y\in \partial\textbf{Y}_{seg}} \left \| p-y \right \|_2 ,  \\
	&\sup_{y\in \partial\textbf{Y}_{seg}} \inf_{p\in \partial\textbf{P}_{seg}} \left \| p-y \right \|_2 \}
	\end{aligned}
	\end{equation}
	where $\partial\textbf{P}_{seg}$ and $\partial\textbf{Y}_{seg}$ represents the surface point sets of $\textbf{P}_{seg}$ and $\textbf{Y}_{seg}$, respectively.
	A large Dice coefficient or a small Hausdorff distance indicates a more accurate segmentation result.

\subsection{Results}

\begin{figure*}[t]
	\begin{center}
		\includegraphics[width=1.0\linewidth]{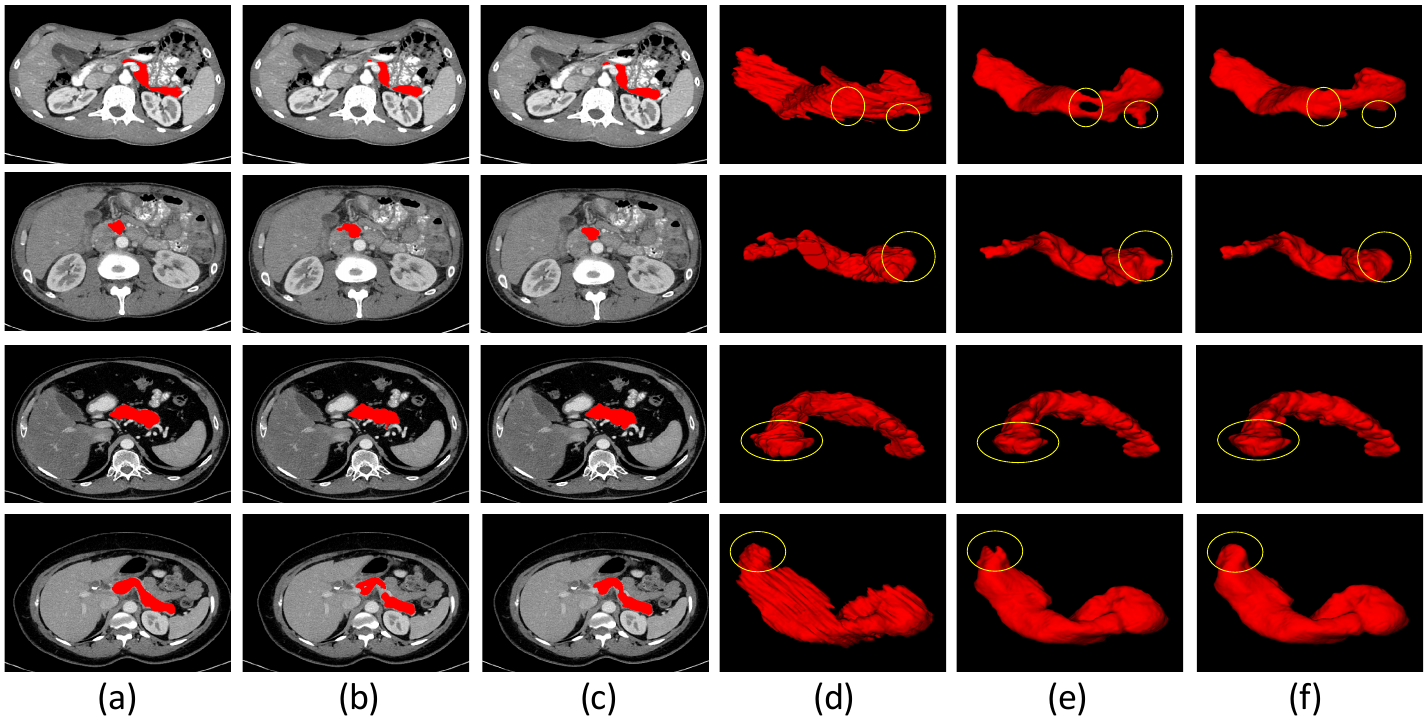}
	\end{center}
	\caption{Comparison of the segmentation results produced by the baseline model and the proposed ConResNet. The left three columns show (a) the ground truth, (b) results of baseline, and (c) results of ConResNet overlaid on CT slices. The right three columns show the 3D visualization of (d) the ground truth, (e) results of baseline, and (f) results of ConResNet. Note that the pancreas is highlighted in red.}
	\label{fig:vis_pancreas}
\end{figure*}

\noindent{\textbf{Comparison on the BraTS dataset:}}
On the BraTS dataset, we compared the proposed ConResNet to six segmentation models top-ranked in the BraTS 2018 Challenge, including CascadeNet \cite{wang2018automatic}, DMFNet \cite{chen2019dmfnet}, OM-Net \cite{zhou2020oneTIP}, DeepSCAN \cite{mckinley2018ensembles}, VAE-Seg \cite{myronenko20183d}, EnsembleNets \cite{zhou2018learning}, nnUnet \cite{isensee2018no}. 
CascadeNet \cite{wang2018automatic} is a triple cascaded network, which hierarchically segments WT, TC and ET. 
OM-Net~\cite{zhou2020oneTIP} separates multi-class segmentation tasks into one deep model and only requires one-pass computation for brain tumor segmentation. 
DMFNet \cite{chen2019dmfnet} is built upon a multi-fiber unit, embedding the weighted 3D dilated convolution to gain multi-scale image representation for improved segmentation.
DeepSCAN \cite{mckinley2018ensembles} is a densely connected segmentation model that uses dilated convolutions to increase the receptive field. 
VAE-seg \cite{myronenko20183d} contains a variational auto-encoder to reconstruct the input image itself in order to regularize the segmentation tasks. 
EnsembleNets \cite{zhou2018learning} is an ensemble of multiple segmentation models, each being specially designed to learn contextual and attentive information. 
nnUnet \cite{isensee2018no} incorporates with additional tricks into UNet to boost the segmentation performance.

The Dice scores and Hausdorff distance obtained by using these models to segment three sub-regions, i.e., ET, WT, and TC, were given in TABLE~\ref{tab:brats}. Note that the Dice scores and Hausdorff distances of compared methods were reported in the original papers. It shows that our ConResNet achieves the largest Dice score for ET segmentation and smallest Hausdorff distance for WT segmentation, whereas nnUnet achieves the largest Dice score for WT segmentation and smallest Hausdorff distance for ET segmentation. Since no model performs consistently better than others when measured by both metrics on three segmentation tasks, we also give the rank (immediately after the metric score) of each model in terms of each metric on each task, where `1' means performing best and `7' means performing worst. To assess the overall performance of each model, we shows the summation of the ranks over all metrics in the last column, which is treated as the golden performance indicators for brain tumor segmentation. It reveals that our ConResNet outperforms other six state-of-the-art methods on this 3D medical image segmentation problem, as indicated by the smallest Sum-Score. It should be noted that, although achieving an Sum-Score close to ours, nnUnet uses additional data for co-training. 
Note that all competing methods use the ensemble strategy. nnUnet and VAE-Seg improve the segmentation performance partly due to the ensemble of five and 10 models, respectively. EnsembleNets averages the probabilities predicted by multiple models, including MC-Net, OM-Net and their variants \cite{zhou2018learning}. Thanks to the context residual learning strategy, we achieve competitive results using the ensemble of three ConResNets, which were trained independently.

\begin{table*}[!t]
\caption{Performance of our ConResNet with / without the context residual mapping and context attention mapping on the BraTS validation dataset.}
\begin{center}
	\begin{tabular}{p{1.8cm}<\centering|c|c|p{1.0cm}<\centering|p{1.0cm}<\centering|p{1.0cm}<\centering|p{1.0cm}<\centering|p{1.0cm}<\centering|p{1.0cm}<\centering}
		\hline
		\multirow{2}{*}{ } & \multirow{2}{*}{Context residual mapping} & \multirow{2}{*}{Context  attention mapping} & \multicolumn{3}{c|}{Dice score (\%)} & \multicolumn{3}{c}{Hausdorff distance} \\ \cline{4-9} 
		Models &                                    &                                     & ET        & WT       & TC       & ET          & WT          & TC          \\ \hline \hline
		Baseline  & $\times$ & $\times$ & 80.66 & 89.15 & 85.89 & 4.46 & 5.23 & 6.30 \\ \hline
		ConResNet  & \checkmark  &  $\times$ & 81.15     & 89.72    & 87.63    & 3.39        & 5.19        & 5.40        \\ \hline
		ConResNet  & \checkmark  & \checkmark & \textbf{81.87}     & \textbf{89.77}    & \textbf{88.14}    & \textbf{3.14}        & \textbf{5.08}        & \textbf{5.16}        \\ \hline
	\end{tabular}
\end{center}
\label{tab:ablation}
\end{table*}

\begin{figure*}[!t]
	\begin{center}
		\includegraphics[width=1.0\linewidth]{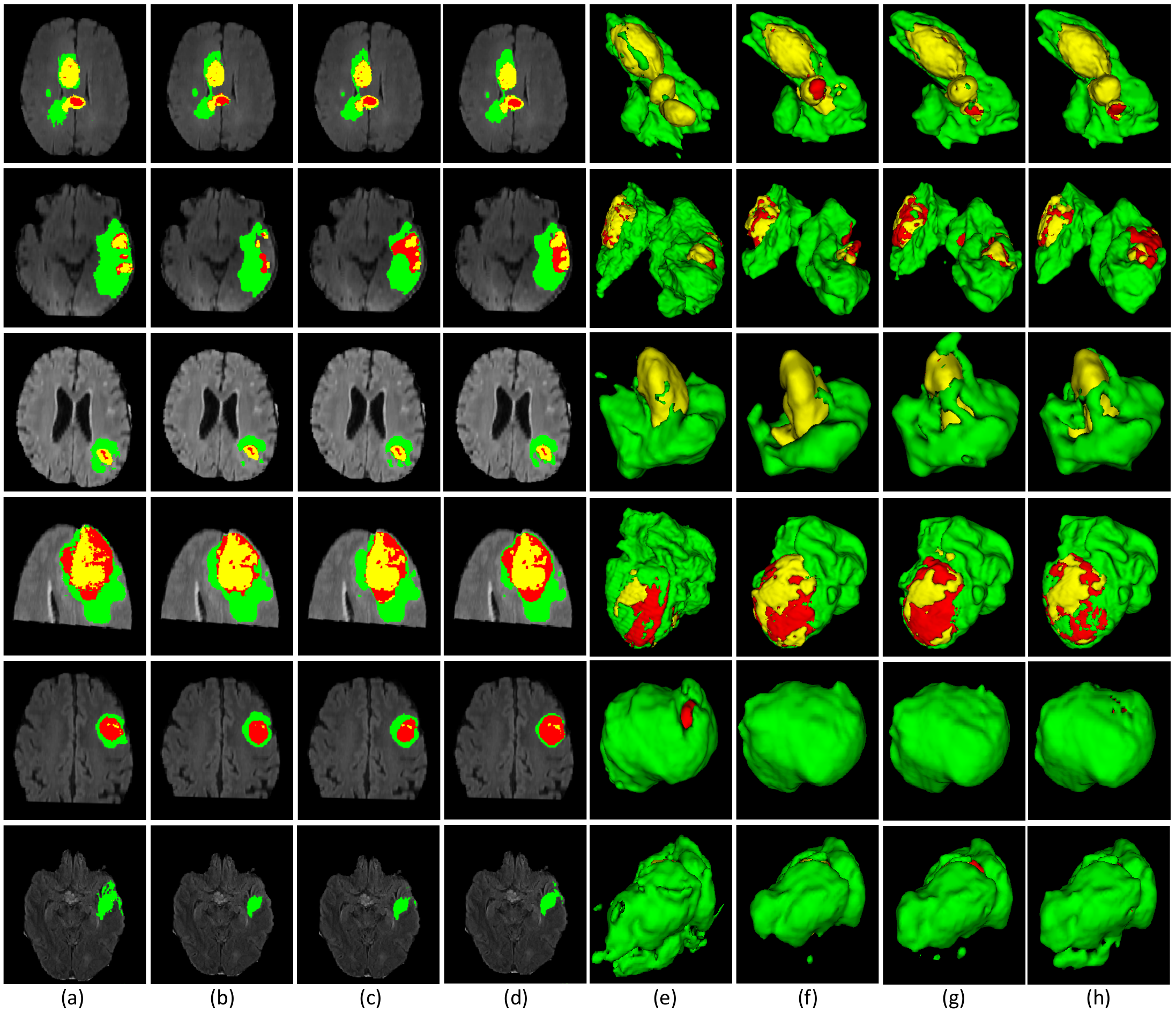}
	\end{center}
	\caption{Segmentation results obtained by applying our ConResNet with / without the context residual mapping and context attention mapping to six cases on the BraTS validation dataset. From left to right: (a) 2D ground truth overlaid on FLAIR slices, (e) ground truth of 3D tumor, and the 2D and 3D segmentation results of (b,f) the baseline model, (c,g) the baseline with context residual, and (d,h) the baseline with both context residual and context attention (ConResNet). ET: Yellow; TC: Yellow + Red; WT: Yellow + Red + Green.}
	\label{fig:visual}
\end{figure*}

\noindent{\textbf{Comparison on the Pancreas-CT dataset:}}
On the Pancreas-CT dataset, we compared the proposed ConResNet to HNN \cite{roth2018spatial}, Attention-UNet \cite{schlemper2019attentionUnet}, CNN-RNN \cite{cai2018pancreas}, RSTN \cite{yu2018recurrent}, ResDSN \cite{zhu20173d}, Bayesian \cite{ma2018novel},  and Coarse2Fine \cite{zhao2019fully}, which are top-ranked segmentation methods on this task.
HNN \cite{roth2018spatial} is a holistically-nested DCNN approach to pancreas localization and segmentation, exploiting multi-view spatial pooling and combining interior and boundary mid-level cues. 
As an improved version of UNet, Attention-UNet \cite{schlemper2019attentionUnet} incorporates a novel attention gate module into the model to force it to focus on target structures of varying shapes and sizes. 
CNN-RNN \cite{cai2018pancreas} employs the recurrent neural networks to address the problem of spatial non-smoothness of inter-slice pancreas segmentation along adjacent slices. 
RSTN \cite{yu2018recurrent} repeatedly transforms the segmentation probability map from previous iterations into the spatial prior and use it in the current iteration to relate the coarse and fine stages. 
ResDSN \cite{zhu20173d} is composed of a coarse model and a fine model, the former obtains the rough location of the pancreas and the latter refines the segmentation based on the coarse result. 
Bayesian \cite{ma2018novel} leverages the Bayesian model to incorporate the rich shape priors learned from statistical shape models into the deep neural network and thus improves the performance in pancreas segmentation. 
Coarse2Fine \cite{zhao2019fully} is also a two stage framework for pancreas segmentation, which contains the coarse segmentation for candidate region generation and the fine segmentation of smaller regions-of-interest.

The mean Dice, max Dice, and min Dice of our ConResNet and these methods were listed in TABLE~\ref{tab:PancreasCT}. Note that the scores of competing methods were adopted from the original papers. It shows ConResNet achieves the highest mean Dice of 86.06\%, highest max Dice of 92.00\%, and highest min Dice of 73.40\%, outperforming the baseline and other competitive methods in terms of all metrics. 

Besides, we visualize the segmentation results of the baseline model that does not contain the ConRes decoder and our ConResNet model in Fig.~\ref{fig:vis_pancreas}. Note that the baseline model has the same encoder and segmentation decoder as ConResNet, including number of channels, network depth, and training strategies. 
It shows that our ConResNet generates better segmentation results, which are more similar to the ground truth than the results of the baseline model. 

\begin{figure}[t]
\begin{center}
	\includegraphics[width=1.0\linewidth]{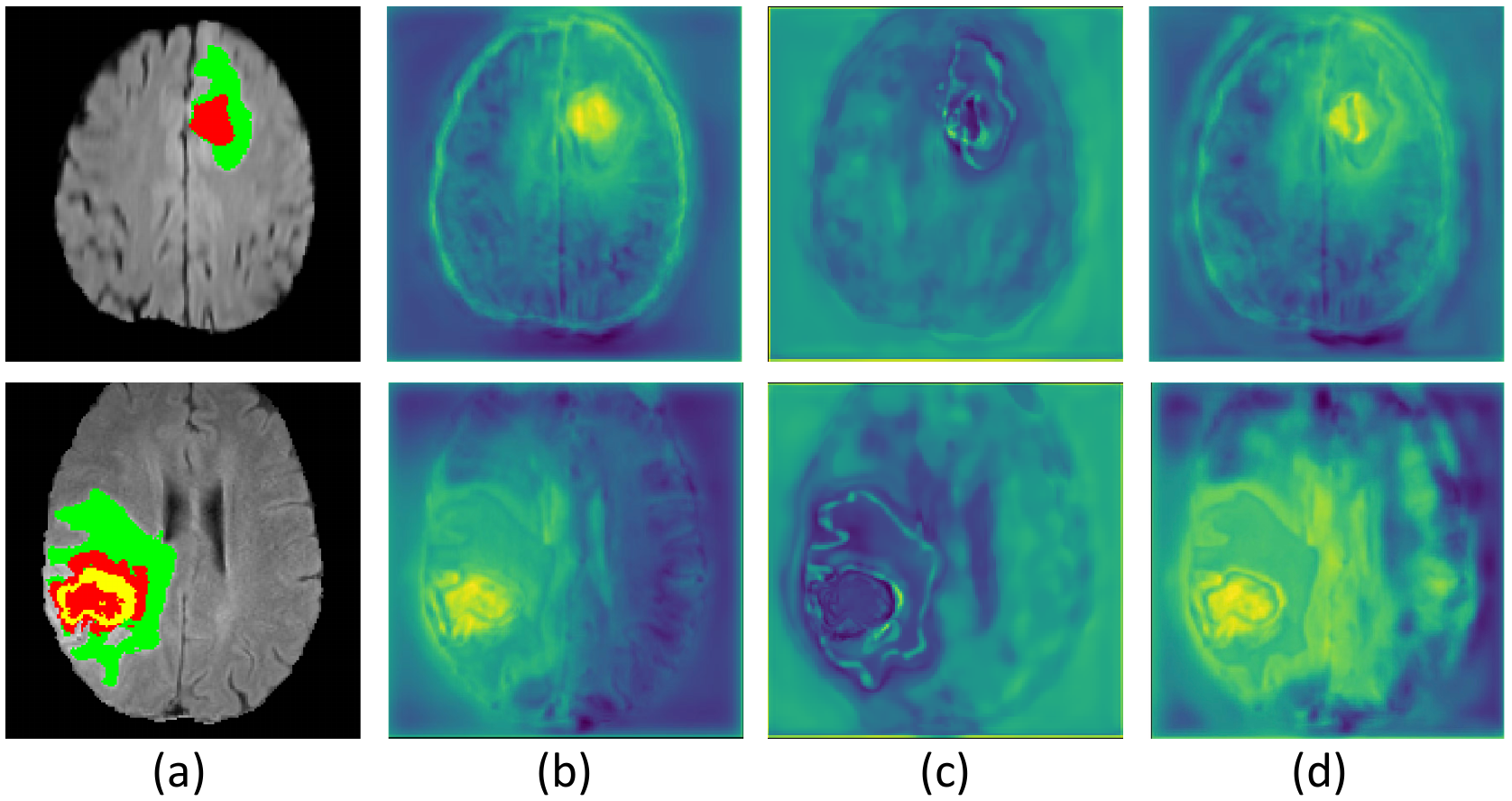}
\end{center}
\caption{Visualization of the learned context residual attention maps: (a) input FLAIR MR slices, (b) feature maps obtained without the attention, (c) learned context residual attention maps, and (d) feature maps obtained with the attention. Note that the feature maps shown here are the channel-wise summation of the output of the last context residual module.}
\label{fig:att2}
\end{figure}

\noindent{\textbf{Visualization of the learned context residual attention maps:}}
In Fig.~\ref{fig:att2}, we visualized the learned context residual attention maps and compared the feature maps obtained using or without using the context attention mapping. It reveals that the learned context residual attention maps highlight the context residual positions of brain tumor sub-regions. With the context attention mapping, the obtained feature maps have an improved ability to highlight the boundaries of those sub-regions, which is beneficial for better segmentation.

\begin{figure}[t]
\begin{center}
	\includegraphics[width=0.8\linewidth]{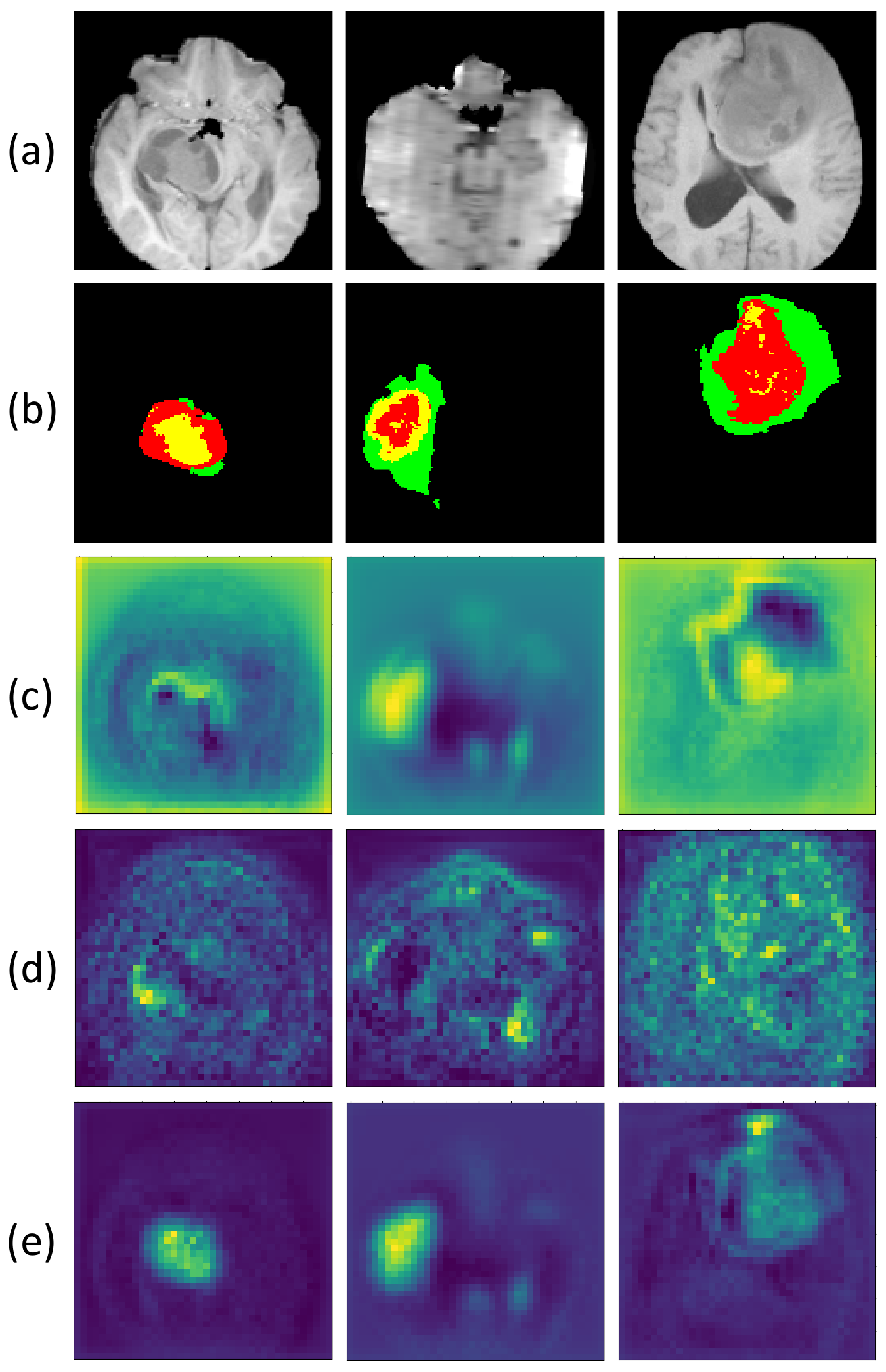}
\end{center}
\caption{Visualization of feature maps produced by different models. All the feature maps are extracted from the output of first stage in the decoder. From top to down: (a) T1-weighted brain MR slices as input, (b) ground truth, and feature maps produced by (c) the baseline model, (d) baseline with context residual, and (e) baseline with both context residual and context attention (ConResNet). Note that the feature maps shown here are the channel-wise summation. }
\label{fig:att}
\end{figure}

\section{Discussion}

\subsection{Ablation Analysis}
The major contribution of this work is to add the ConRes decoder and context residual models, which enable both context residual mapping and context attention mapping, to an encoder-decoder structure for improved segmentation performance. To verify the effectiveness of this design, we performed ablation experiments on the BraTS dataset.
For the convenience of quantitative evaluation, we randomly selected 35 cases from training set to form a local validation set. Thus, we have 250 training cases when evaluating on the validation set (35 cases) and have 285 cases when evaluating on the testing set (66 cases). The performance our ConResNet with / without the context residual mapping and context attention mapping on the local validation set was listed in Table ~\ref{tab:ablation}. Note that we kept other settings such as the network width, depth, and training strategies the same to ensure a fair comparison.

It reveals that using the context residual mapping helps the segmentation model achieve larger Dice scores and smaller Hausdorff distances on ET, WT, and TC, and the performance improvement is substantial in all metrics. Meanwhile, incorporating the context attention mapping into the model with the context residual mapping can further improve the segmentation performance in all metrics. Consequently, comparing to the baseline model that uses neither the context residual mapping nor the context attention mapping, our ConResNet improves the Dice scores by 1.66\%, 0.83\%, and 2.27\% and decreases the Hausdorff distance by 1.76, 0.41, and 1.44 for the segmentation of ET, WT, and TC, respectively.

The segmentation results overlaid on FLAIR slices and the 3D visualization of segmented brain tumors were displayed in Fig.~\ref{fig:visual}. It shows that the results produced by our ConResNet are more similar to the ground truth. Both the quantitative evaluation in Table~\ref{tab:ablation} and qualitative comparison in Fig.~\ref{fig:visual} demonstrate the effectiveness of the proposed ConResNet for 3D medical image segmentation.

In Fig.~\ref{fig:att}, we also visualize the channel-wise summation of the feature maps obtained by the baseline model, the model with only context residual mapping, and our ConResNet that uses both context residual mapping and context attention mapping. It shows that the feature maps produced by our model focus more on the target region than the feature maps produced by other models, which is beneficial for the segmentation.

\subsection{Deep Supervision}
\label{deepsupervision}
\begin{figure}[t]
	\begin{center}
		\includegraphics[width=0.99\linewidth]{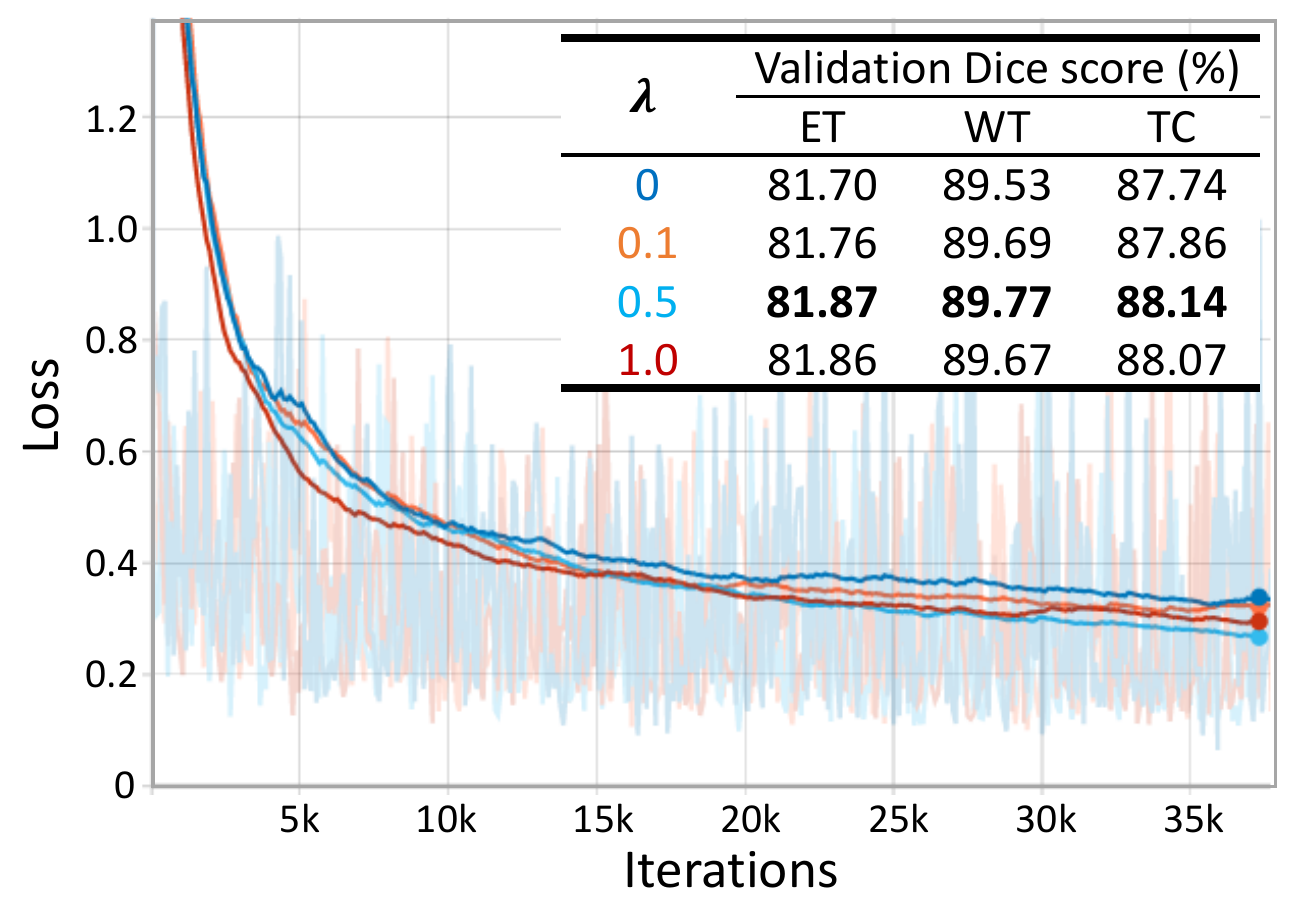}
	\end{center}
	\caption{Plot of training loss of our ConResNet versus iterations when setting $\lambda$ to different values. The table reports the validation performance under the different setting of $\lambda$. }
	\label{fig:lambda}
\end{figure}

The proposed ConResNet also uses deep supervision, which is controlled by the trade-off parameter $\lambda$ (see Eq.(8)). To validate the contribution made by deep supervision, we trained the model on the BraTS dataset four times when setting $\lambda$ to 0, 0.1, 0.5, and 1, respectively. Note that $\lambda=0$ means no deep supervision. In Fig.~\ref{fig:lambda}, we observe the lowest training loss and the best performance on the local validation set when $\lambda = 0.5$. Therefore, we empirically set $\lambda$ to 0.5 in our experiments.

\subsection{Learning Context Residual along Other Views}
Our ConResNet is not limited by the inter-slice residual information along the axial view. It is easy to employ the proposed context residual learning to sagittal and coronal views. Taking the sagittal view for example, we compute the position-wise absolute difference between each pair of adjacent features along longitudinal axis, and train our ConResNet to predict the sagittal-view context residual mask. As shown in TABLE~\ref{tab:views}, we compared the performance of our ConResNet model on the local validation set, considering the residual information along the axial view, sagittal view, and coronal view, respectively. It shows that comparable performance was obtained when considering each of three views. It also reveals that ConResNet with any view is superior to the baseline model in terms of all evaluation metrics. These results demonstrate the effectiveness of the proposed context residual learning, which is not limited to the axial view.

\begin{table}[!t]
\caption{Performance of ConResNet when learning context residual along axial, sagittal, and coronal views.}
\begin{center}
\begin{tabular}{c|c|c|c|c|c|c|c}
\hline
\multirow{2}{*}{Method} & \multirow{2}{*}{View} & \multicolumn{3}{c|}{Dice}                        & \multicolumn{3}{c}{Hausdorff distance}       \\ \cline{3-8} 
 & & ET & WT & TC & ET & WT & TC \\ \hline \hline
Baseline & - & 80.66 & 89.15 & 85.89 & 4.46 & 5.23 & 6.3 \\ \hline
ConResNet & Axial & \textbf{81.87} & 89.77  & 88.14 & 3.14 & 5.08 & 5.16 \\ \hline
ConResNet & Sagittal & 81.33 & \textbf{89.81} & 87.78 & 2.98 & \textbf{4.51} & 5.43  \\ \hline
ConResNet & Coronal & 81.51 & 89.70 & \textbf{88.20} & \textbf{2.84} & 4.76 & \textbf{4.63} \\ \hline
\end{tabular}
\end{center}
\label{tab:views}
\end{table}

\subsection{Applying to Multi-class Problems}
Let us consider a multi-class segmentation task with $C$ categories, where the ground-truth segmentation mask is denoted by $\textbf{Y}_{seg} \in \textbf{R}^{(C+1)\times S\times H \times W}$. For each category $c\in \{1,2,...,C\}$, we specifically compute its context residual mask as 
\begin{equation}
\textbf{Y}_{res}^{c, s+1, h, w} = \left |  \textbf{Y}_{seg}^{c, s+1, h, w} - \textbf{Y}_{seg}^{c, s, h, w} \right |
\end{equation}
In the multi-class mode, ConResNet predicts the segmentation mask after the $softmax$ activation and predicts the residual mask of each category after the $sigmoid$ activation. Besides, we ignore the residual mask of background and just consider the error back-propagation of C categories during the training process. We compared the multi-class ConResNet to its binary counterpart on the BraTS dataset. The brain tumor segmentation task is formulated either into three binary segmentation problems, including ET vs. others, WT vs. others, and TC vs. others, or into a four-class segmentation problem, where the four classes include: 0 - background, 1 - necrotic and non-enhancing tumor core, 2 - peritumoral edema, and 3 - enhancing tumor. We still evaluated the segmentation performance on three sub-regions as done in the BraTS Challenge. TABLE~\ref{tab:multiclass} shows the performance of the baseline model and our ConResNet in either the binary or multi-class mode on the local validation set. It reveals that the binary mode leads to the better performance than the multi-class mode, which may owe to the direct optimization of three sub-regions. However, our ConResNet achieves better performance than the baseline model in both binary and multi-class settings.

\begin{table}[!t]
\caption{Comparison of ConResNet and the baseline model in  binary or multi-class mode. Mode: B - ``Binary"; M - ``Multi-class".}
\begin{center}
\begin{tabular}{c|c|c|p{0.5cm}<\centering|p{0.5cm}<\centering|p{0.5cm}<\centering|p{0.4cm}<\centering|p{0.4cm}<\centering|p{0.4cm}<\centering}
\hline
\multirow{2}{*}{Method} & \multicolumn{2}{c|}{Mode} & \multicolumn{3}{c|}{Dice}                        & \multicolumn{3}{c}{Hausdorff distance}       \\ \cline{2-9} 
 & B & M & ET & WT & TC & ET & WT & TC \\ \hline \hline
Baseline                & \checkmark &                  & 80.66          & 89.15          & 85.89          & 4.46          & 5.23          & 6.30           \\ \hline
ConResNet               & \checkmark &                  & \textbf{81.87} & \textbf{89.77} & \textbf{88.14} & \textbf{3.14} & \textbf{5.08} & \textbf{5.16} \\ \hline \hline
Baseline                &             & \checkmark & 79.85          & 88.87          & 85.74          & 3.48          & 5.73          & 5.42          \\ \hline
ConResNet               &             & \checkmark & \textbf{81.04} & \textbf{89.12} & \textbf{86.51} & \textbf{3.13} & \textbf{4.90}  & \textbf{5.37} \\ \hline
\end{tabular}
\end{center}
\label{tab:multiclass}
\end{table}

\subsection{Comparing to Boundary Loss-based Methods}
The methods reported in \cite{karimi2019hausdorff} and \cite{kervadec2018boundary} use newly designed boundary-wise loss functions to force the model to pay attention to boundary pixels on the object surface. Different intrinsically from them, the proposed ConResNet aims to learn the inter-slice context residual, which contains the essential and intriguing morphological information of the tumor, and use the context residual to boost the segmentation performance via simultaneous context residual mapping and context attention mapping. Although context residual voxels appear on or near to the tumor surface, using the proposed context residual learning is different from defining a boundary loss. In fact, the boundary loss can be incorporated into our model to possibly further improve its performance. In this case, our ConResNet has a compound loss function
\begin{equation}
L_{compound} = \alpha \times L + (1-\alpha) \times L_{BD}
\end{equation}
where $\alpha = 1- t \cdot \frac{1-0.01}{T}$ is a weighting factor, and $L_{BD}$ is the boundary loss proposed in \cite{kervadec2018boundary}. We compared the baseline model and our ConResNet model with and without using the boundary loss on the local validation set. 
TABLE~\ref{tab:boundaryloss} show that using the boundary loss can further improve the performance of the baseline model and our ConResNet. Nevertheless, it also shows that, although the boundary loss is beneficial for this segmentation task, our ConResNet even outperforms the baseline model with the boundary loss.
In general, adding a boundary loss or using the context residual learning is able to effectively help the model pay more attention to boundaries so as to improve the segmentation performance, particularly in localizing the surface of each target volume. Interestingly, both approaches address the discrepancies in the boundaries from different perspectives. Jointly using the context residual learning and a boundary loss is able to produce a mutual promotion.

\begin{table}[!t]
\caption{Comparison of baseline and ConResNet with and without boundary loss (BD).}
\begin{center}
\begin{tabular}{l|c|c|c|c|c|c}
\hline
\multirow{2}{*}{Method} & \multicolumn{3}{c|}{Dice score} & \multicolumn{3}{c}{Hausdorff distance} \\ \cline{2-7} 
 & ET & WT & TC & ET & WT & TC \\ \hline \hline
Baseline & 80.66 & 89.15 & 85.89 & 4.46 & 5.23 & 6.3 \\ \hline
Baseline+BD & \textbf{81.13} & \textbf{89.31} & \textbf{86.49} & \textbf{4.03} & \textbf{5.19} & \textbf{6.19} \\ \hline \hline
ConResNet & 81.87 & 89.77 & 88.14 & 3.14 & 5.08 & 5.16 \\ \hline
ConResNet+BD & \textbf{81.90}  & \textbf{89.87} & \textbf{88.29} & \textbf{2.72} & \textbf{4.53} & \textbf{4.55} \\ \hline
\end{tabular}
\end{center}
\label{tab:boundaryloss}
\end{table}

\subsection{Efficiency Analysis}
Our ConResNet has dual decoders, which, inevitably, raises the number of parameters and computations. 
TABLE~\ref{tab:param} lists the number of parameters, model size, Giga Floating-point Operations Per Second (GFLOPs) of VAE-Seg, baseline and our ConResNet.
Note that the GFLOPs was calculated when the input is a 3D volume of size $80\times160\times160$.
It shows that our ConResNet suffers from slightly more parameters (an increase of 2.66\%), more computations (an increase of 18.64\%), and a larger mode size (an increase of 2.53\%) than the baseline. However, considering the performance improvement shown in TABLE~\ref{tab:ablation} and Fig.~\ref{fig:visual}, we believe that such a moderate increase of the complexity is acceptable. Moreover, although it is somewhat slower than the baseline model, our ConResNet can perform brain tumor segmentation almost in real-time, and hence has the potential to be used in clinical practice.

\begin{table}[!t]
\caption{Number of parameters, model size and GFLOPs of different models.}
\begin{center}
\begin{tabular}{c|c|c|c}
\hline
Models    & \begin{tabular}[c]{@{}c@{}}\#Params \\ ($\times 10^6$)\end{tabular} & \begin{tabular}[c]{@{}c@{}}Model size \\ (MB)\end{tabular} & GFLOPs \\ \hline
VAE-Seg \cite{myronenko20183d} & 23.6 & 89.7 & 856.1 \\ \hline
Baseline  & 18.8 & 75.2 & 453.9 \\ \hline
ConResNet & 19.3  & 77.1 & 538.5 \\ \hline
\end{tabular}
\end{center}
\label{tab:param}
\end{table}

\section{Conclusion}
In this paper, we propose the ConResNet with explicit 3D context learning to boost the ability of DCNNs to perceive inter-slice context for accurate segmentation of volumetric medical images. We have evaluated this model on the BraTS dataset and Pancreas-CT dataset. Our results indicate that the proposed ConResNet outperforms state-of-the-art method on both brain tumor segmentation and pancreas segmentation tasks. Our ablation study also demonstrates the effectiveness of the proposed context learning, including the context residual mapping and context attention mapping.
In our future work, we plan to treat the predicted residual mask as a prior that highlights the error-prone regions and to concatenate it with the input image as the input of another segmentation network for refined result.
Besides, we will investigate how to combine the proposed 3D context learning with self-supervised learning, and thus extend this work to semi-supervised segmentation problems.


%

\ifCLASSOPTIONcaptionsoff
  \newpage
\fi

\bibliographystyle{IEEEtran}
\bibliography{bibtex/IEEEexample}

\end{document}